\input epsf
\input cp-aa
\overfullrule 0pt
\newcount\eqnumber\eqnumber=1
\def\new{{\the\eqnumber}\global\advance\eqnumber by 1}
\def\eqnam#1{\xdef#1{\the\eqnumber}}
\newcount\fignumber\fignumber=1
\def\nfig{\global\advance\fignumber by 1}
\def\fignam#1{\xdef#1{\the\fignumber}}
\def\nompsy#1{\vbox{\epsfbox{#1}\vfil}} 
\fignam{\FBAS}     \nfig
\fignam{\FLS}      \nfig
\fignam{\FBAL}     \nfig
\fignam{\FBASFRT}  \nfig
\fignam{\FBAZT}    \nfig
\fignam{\FBASFRTW} \nfig

\MAINTITLE{Bar strength and star formation activity in late-type barred
galaxies}

\AUTHOR{L.~Martinet and D.~Friedli
  \PRESADD{D\'epartement de physique, Universit\'e Laval, Ste-Foy,
           QC, G1K~7P4, Canada; dfriedli\at phy.ulaval.ca}
}
\INSTITUTE{Observatoire de Gen\`eve, CH-1290 Sauverny, Switzerland

\noindent
E-mail: Louis.Martinet\at obs.unige.ch
}

\OFFPRINTS{L.~Martinet}

\DATE{Received 23 September 1996 / Accepted 24 December 1996}

\ABSTRACT{
With the prime aim of better probing and understanding the intimate
link between star formation activity and the presence of bars, a
representative sample of 32 non-interacting late-type galaxies with
well-determined bar properties has been selected.  We show that all
the galaxies displaying the highest current star forming activity have
both strong and long bars. Conversely not all strong and long bars are
intensively creating stars.  Except for two cases, strong bars are in
fact long as well.  Numerical simulations allow to understand these
observational facts as well as the connection between bar axis ratio,
star formation activity, and chemical abundance gradient: Very young
strong bars are first characterized by a raging episode of star
formation and two different radial gaseous abundance gradients, one
steep in the bar and one shallow in the disc.  Then, principally due
to gas consumption, galaxies progressively fall back in a more
quiescent state with a nearly flat abundance gradient across the whole
galaxy.  On the contrary, weak bars are unable to trigger significant
star formation or to generate flat abundance gradients.  The selected
galaxies have tentatively been classified in four classes
corresponding to main stages of secular evolution scenario. }

\THESAURUS{4(11.01.1; 11.05.2; 11.19.3; 13.09.1)}
\maketitle

\MAINTITLERUNNINGHEAD{Bar strength and star formation activity}

\titlea{Introduction}
Many clues of secular dynamical evolution in disc galaxies have
recently been detected by observations or suggested by numerical
simulations and theoretical approaches. The most significant
progresses in the field has come from considerations of facts
neglected or not well understood in the past. For instance, the role
of gas, the effects of interactions not only between galaxies but also
between various components of a given system, the necessity to fully
take into account 3D structures, the interplay between star formation
and dynamical mechanisms, etc.  In fact, discs are the seat of
evolutionary processes on timescales of the order of the Hubble time
or less (see the reviews by Kormendy 1982; Martinet 1995; Pfenniger
1996; see also e.g. Pfenniger \& Norman 1990; Friedli \& Benz 1993,
1995; Courteau et al. 1996; Norman et al. 1996). In particular, bars
do play a decisive role in such processes.

The presence of non-axisymmetric components seems to be a necessary
condition for the onset of nuclear activity (Moles et al. 1995).
However, the precise link between the presence of stellar bars and the
star formation activity is not well established and even somewhat
controversial (see e.g. Hawarden et al. 1996).  Some authors claimed
that star formation is enhanced in barred galaxies (Hawarden et
al. 1986; Dressel 1988; Arsenault 1989; Huang et al. 1996), whereas
others suggested that barred galaxies have star formation levels
similar or lower than those in normal spirals (Pompea \& Rieke 1990;
Isobe \& Feigelson 1992).  For instance, according to Hawarden et
al. (1986), more than one third of SB galaxies have 25~$\mu$m excess
attributed to vigorous star formation in a circumnuclear ring located
near the inner Lindblad resonance. However, this result generates many
unanswered questions: Why do the other galaxies show no significant
excess of star formation activity?  Is the star formation enhancement
dependent on the Hubble type as well?  Is the 25~$\mu$m excess a
reliable indicator of star formation? Where are the preferential sites
of star formation located?  Etc.

The aim of the present paper is threefold: 1) To quantitatively
confirm through observational data from the literature that the
strength and/or the length of a bar is a decisive factor for enhancing
star formation activity as suggested by numerical simulations
(e.g. Friedli \& Benz 1993, 1995), observations (Martin 1995), as well
as by our preliminary study (Friedli \& Martinet 1996). 2) To
re-discuss the influence of the bar strength and the star formation
efficiency on the radial chemical gradient in the continuation of the
work by Martin \& Roy (1994). 3) To suggest steps of evolution in
barred galaxies taking into account the various points previously
mentioned.

This paper is structured as follows: In Sect.~2, we discuss various
indicators and estimators of star formation usually considered in the
literature, whereas the selection of galaxies used in this study is
given in Sect.~3. Section~4 is devoted to the presentation of the
various links found in our sample between observed quantities, like
bar axis ratio, bar length and star formation activity as well as by
considering a simple theoretical model able to link the radial
abundance gradient, the bar strength, and the star formation
efficiency.  In Sect.~5 some other jigsaw pieces have been inserted
thanks to a new set of numerical simulations. In Sect.~6, we discuss
and put the previous results in the general frame of secular evolution
of disc galaxies, and finally we summarize our main conclusions in
Sect.~7.

\titlea{Star formation activity}
\titleb{Star formation tracers}
There does not seem to exist any observational quantitative estimator
of active star formation devoid of ambiguity.  The present situation
is briefly discussed and summarized in the following four subsections,
whereas our final choice is discussed in the fifth one.

\titlec{H$\alpha$}
One can expect that H$\alpha$ emission leads to a good estimator of
instantaneous star formation rate (SFR) insofar as it can be checked
that they have been corrected for extinction, for instance by
comparing with thermal radio luminosities (Sauvage \& Thuan 1992).
The reader is referred to Kennicutt (1983), Kennicutt \& Kent (1983),
Keel (1983), and Pogge (1989) for data on the H$\alpha$ emission
properties of normal galaxies.  For a general discussion see Kennicutt
(1989).

\titlec{UV}
UV fluxes in a bandpass of about 125\AA\ centered at 2000\AA\ have
been used by Donas et al. (1987) to obtain quantitative estimates of
the current SFR. Difficulties and uncertainties in the correction for
extinction are similar to those present for the H$\alpha$ data.

\titlec{FIR}
The connection between IRAS far-infrared (FIR) and H$\alpha$ emissions
is still controversial. Sauvage \& Thuan (1992) outline the strong
linearity in a $\log(L_{\rm H\alpha}) - \log(L_{\rm FIR})$
correlation. They show that the decrease of the $L_{\rm FIR}/L_{\rm
H\alpha}$ ratio along the Hubble sequence can be explained by a model
of FIR-emitting ISM consisting of two components, i.e. star forming
regions and quiescent cirrus-like regions as originally introduced by
Lonsdale Persson \& Helou (1987) and Rowan-Robinson \& Crawford
(1989).

Comparing these last approaches, we observe that in two-colour diagram
$\log(S_{60}/S_{100})$ versus $\log(S_{12}/S_{25})$, galaxies with
$\log(S_{60}/S_{100}) \!>\! -0.30$ and $\log(S_{12}/S_{25}) \!<\!
-0.35$ have a probable contribution to the flux from recent star
formation at least of the order of 50\%. Moreover, Sauvage \& Thuan
(1992) infer from the decreasing fraction of $L_{\rm FIR}$ associated
to the cirrus from Sa to Sdm that the high mass star formation
efficiency increases toward late spiral types. This efficiency is
defined as the fraction of FIR-emitting ISM directly associated with
star formation.

Various authors directly used FIR colour indices as estimators of
current star formation activity. According to Puxley et al. (1988)
galaxies with $\log(S_{12}/S_{25}) \!<\! -0.35$ are believed to
contain regions of star formation. Eskridge \& Pogge (1991) admit that
$\log(S_{60}/S_{100}) \approx -0.35$ is the value above which a major
part of the FIR emission is expected to be due to star
formation. Sekiguchi (1987) indicates that the fraction of 60~$\mu$m
emission attributable to a warm component can be used as an indicator
of star formation activity.  Dultzin-Hacyan et al. (1990) consider
that the best IRAS tracer of recent star formation is the ratio
$\log(S_{25}/S_{100})$. In normal galaxies, the mean value of this
ratio is $-1.30$, whereas for liners and starbursts it is respectively
$-1.15$ and $-0.75$.  Moreover, considering $\Delta \rm H\alpha$, the
global to central H$\alpha$ flux ratio taken from Kennicutt (1983) and
Keel (1983), we observe a correlation between $\Delta \rm H\alpha$ and
$\log(S_{25}/S_{100})$. This clearly suggests that some information on
star formation activity is contained in this indicator in spite of the
caveats mentioned above.

From the analysis of these different suggestions, we can admit that
objects for which most of the FIR emission is due to current star
formation are separated from those with FIR colours indistinguishable
from galactic cirrus by the alternative conditions:
$\log(S_{12}/S_{25}) \la -0.35$, $\log(S_{60}/S_{100}) \ga -0.35$,
$\log(S_{25}/S_{100}) \ga -1.15$, or $\log(S_{25}^2/S_{12}S_{100}) \ga
-0.80$.  The ratio $\alpha_2/\alpha_1$ of the respective contributions
of starburst and cirrus in the observed spectrum of IRAS galaxies
(Rowan-Robinson \& Crawford 1989) could also be considered. The
frontier in color indices defined above corresponds to
$\alpha_2/\alpha_1 \approx 1$.

Finally, in view of strengthening our statement, we can compare the
global values of estimators for various galaxies with local values
obtained in nearby galaxies.  In fact, the FIR sources in M31 seem to
coincide with giant HII regions complexes (Xu \& Helou 1996). The same
observation is reported by Rice et al. (1990) from IRAS maps of M33 or
by Xu et al. (1992) for the LMC.  Tomita et al. (1996) localize
galaxies in different part of the $\log(S_{100}/S_{60})$ versus
$L_{\rm FIR}/L_{\rm B}$ diagram corresponding to HII regions, non-HII
regions and central regions of M31. This approach suggests that
$L_{\rm FIR}/L_{\rm B}$ could be a useful indicator of current versus
recent star formation rates.

\titlec{Radio}
The origin of the tight correlation between the FIR and the radio
continuum emission of late-type galaxies has been discussed by various
authors (e.g. Helou 1991 for a review). The 60~$\mu$m to 20~cm
IR-to-radio ratio seems to be a signature of star formation activity
resulting from stars with $m_* \ga 5 \, \rm M_\odot$. As shown by
Puxley et al. (1988), the majority of SB galaxies with
$\log(S_{12}/S_{25}) \!<\! -0.35$ have central radio source emission
arising from the center of a burst of star formation. However, such an
emission could also be powered by a density enhancement or a darkened
active nucleus.

\titlec{Our choice}
The necessity to have a sample of galaxies as large as possible with
both data on bar morphology and star formation activity leads us to
prefer FIR to H$\alpha$ or radio data.  We also have to take into
account all the different caveats and problems listed in Sect.~2.1.3.
Therefore, after having consulted various sources of data, we choose
to use $\log(S_{25}/S_{100})$ as indicator of star formation activity
for the late-type SBs galaxies from Martin's catalogue (Martin 1995).
The effects observed are qualitatively confirmed by using other color
indices.

\titleb{Star formation rates}
For the sake of comparison with numerical simulations (see Sect.~5),
we would like to have estimates of the SFRs. However, the SFR values
calculated from $L_{\rm FIR}$ data must be used with caution. The
results are strongly affected by at least three factors: 1) The choice
of the shape and the mass range for the IMF. 2) The IR wavelength
range used in published data. 3) The cirrus contribution to $L_{\rm
FIR}$ as a function of Hubble type.

Concerning the IMF, it must be emphasized that in the general relation
${\rm SFR_{FIR}} \!=\! k L_{\rm FIR}$, the value of $k$ will strongly
vary depending on the IMF. For instance, for a Salpeter IMF with mass
ranges 0.1 -- 60 M$_\odot$, 2 -- 60 M$_\odot$, 8 -- 60 M$_\odot$, $k$
will respectively be $6 \cdot 10^{-10}$, $3 \cdot 10^{-10}$, and $1
\cdot 10^{-10}$ (Telesco 1988). The reality of a top-heavy IMF for
starburst galaxies is still debated (see e.g. Sommer-Larsen 1996).

The main source of data for $L_{\rm FIR}$ still come from IRAS fluxes
which do not cover the whole IR spectrum. Some extrapolations have
been tried to correct for missed fluxes beyond 120~$\mu$m and
shortward 40~$\mu$m. This correction depends on the ratio
$S_{60}/S_{100}$ (see e.g. Young et al. 1989).

Finally, the percentage coming from the cirrus contribution can be
accounted for by using the two-component model introduced by Lonsdale
Persson \& Helou (1987).  Following Sauvage \& Thuan (1994), a
contribution of 77\% for Sbc's, 70\% for Sc's, and 45\% for Scd's is
adopted.

The previous considerations mean that only relative SFRs are really
relevant although access to absolute SFRs remains necessary. Assuming
the same IMF for all galaxies in the sample and a linear relation
between the FIR flux and the SFR, we use the following relation
inferred from Telesco (1988):
$$
{\rm SFR_{\rm FIR} [M_\odot \, yr^{-1}]} = 
6 \cdot 10^{-10} \tilde{L}_{\rm FIR} [\rm L_\odot] \, ,
\eqno(\new)
$$
where $\tilde{L}_{\rm FIR}$ is $L_{\rm FIR}$ taken on the 1 --
500~$\mu$m range from Young et al. (1989), corrected for the cirrus
contribution according to Sauvage \& Thuan (1994).  Equation~1
concerns the mass range 0.1 -- 60 M$_\odot$. The numerical coefficient
must be divided by two if the mass range is 2 -- 60 M$_\odot$ as
mentioned before.  We introduce a normalizing factor $\rm
\overline{SFR}_{FIR} \!=\! 0.72 \, \rm M_\odot \, yr^{-1}$ obtained by
inserting into Eq.~1 the mean FIR luminosity for normal Sc galaxies,
i.e. $\overline{L}_{\rm FIR} \!=\!  4 \cdot 10^9 \, \rm L_\odot$
(Becklin 1986). Then, we relate all the SFRs to $\rm
\overline{SFR}_{FIR}$,
$$
f \equiv \, {\rm SFR_{FIR}(galaxy) \over \overline{SFR}_{FIR}} 
  = C \, {L_{\rm FIR}({\rm galaxy}) \over \overline{L}_{\rm FIR}} \, ,
\eqno(\new)
$$
where $C$ is a weighting factor to take into account the respective
cirrus contribution ($C \!=\! 0.77$ for Sbc's; $C \!=\! 1.00$ for
Sc's; $C \!=\! 1.83$ for Scd's).  Both $f$ and $\rm SFR_{FIR}$ are
given in Table~1.

For some galaxies in our sample, the SFR deduced from the H$\alpha$
emission, ${\rm SFR_{\rm H\alpha}}$, has been given by Kennicutt
(1983) and is also indicated in Table~1.  As noted by the author,
the individual entries probably possess uncertainties of the order
$\pm 50\%$ due to variable extinction. The ${\rm SFR_{\rm H\alpha}}$
is not directly comparable to $f$. However, for our purpose we can be
satisfied when observing a qualitative agreement on high and low
values of the SFR inferred from FIR and H$\alpha$ data. Using the same
assumptions for the IMF, the distance, and the corrections for
extinction as Kennicutt (1983), the $\rm SFR_{UV}$ (Donas et al. 1987)
are on average larger by a factor 1.2 than the ${\rm SFR_{\rm
H\alpha}}$ for common objects.

\begtabfullwid
\tabcap{1}{Galaxy sample}
\vbox{
\halign{
& # 
&       #\hfill \quad
& \hfill#\hfill \quad
& \hfill#\hfill \quad
& \hfill#\hfill \quad
& \hfill#\hfill \quad
& \hfill#\hfill \quad
& \hfill#\hfill \quad
& \hfill#\hfill \quad
& \hfill#\hfill \quad
& # \cr
\noalign{\hrule\medskip}
& \hfill Names \hfill
& \hfill $(b/a)_i$ \hfill 
& \hfill $2L_i/D_{25}$ \hfill 
& \hfill $\log(S_{25}/S_{100})$ \hfill 
& \hfill $f$ \hfill 
& \hfill $\rm SFR_{FIR}$ \hfill 
& \hfill $\rm SFR_{H_\alpha}$ \hfill 
& \hfill $d[{\rm O/H}]/dR$ \hfill
& \hfill Class &\cr
\noalign{\smallskip}
& \hfill (1) \hfill
& \hfill (2) \hfill 
& \hfill (3) \hfill 
& \hfill (4) \hfill 
& \hfill (5) \hfill 
& \hfill (6) \hfill 
& \hfill (7) \hfill 
& \hfill (8) \hfill 
& \hfill (9) \hfill &\cr
\noalign{\medskip\hrule\medskip}
& NGC~~578 & 0.87 & 0.09 & $-1.31$ $^S$ &    &      &      &                                &       & \cr
& NGC~1313 & 0.63 & 0.12 & $-1.46$ $^R$ &    &      &      &   0.000 $^{W}$                 &       & \cr
& NGC~1365 & 0.51 & 0.27 & $-1.20$ $^R$ &    &      &      & $-0.050$ / 0.000 $^*$ $^{R}$   &       & \cr
& NGC~1637 & 0.52 & 0.13 & $-0.98$ $^S$ & 0.3&  0.2 &  0.9 &                                &  (IV) & \cr
& NGC~1784 & 0.58 & 0.30 & $-1.16$ $^D$ &    &      &      &                                &       & \cr
& NGC~2997 & 0.85 & 0.04 & $-1.31$ $^R$ &    &      &      & $-0.093$ $^{M_1}$              &       & \cr
& NGC~3184 & 0.98 & 0.09 & $-1.25$ $^S$ & 2.2&  1.5 &      & $-0.101$ $^{M_1}$              &    I  & \cr
& NGC~3344 & 0.86 & 0.06 & $-1.33$ $^S$ & 0.6&  0.4 &      & $-0.231$ $^Z$                  &    I  & \cr
& NGC~3359 & 0.32 & 0.20 & $-1.45$ $^S$ & 1.7&  1.2 &      & $-0.070$ / 0.006 $^*$ $^{M_2}$ &   IV  & \cr
& NGC~3486 & 0.74 & 0.07 & $-1.69$ $^S$ & 0.3&  0.2 &  2.2 &                                &    I  & \cr
& NGC~3686 & 0.35 & 0.21 & $-1.35$ $^D$ &    &      &      &                                &       & \cr
& NGC~3726 & 0.70 & 0.07 & $-1.34$ $^S$ & 1.9&  1.3 &  2.2 &                                &    I  & \cr
& NGC~3887 & 0.50 & 0.24 & $-1.43$ $^S$ & 1.2&  0.8 &      &                                &   IV  & \cr
& NGC~3953 & 0.89 & 0.17 & $-1.40$ $^S$ & 3.3&  2.3 &      &                                &   (I) & \cr
& NGC~3992 & 0.58 & 0.30 & $-1.37$ $^Y$ & 2.0&  1.4 &      &                                &   IV  & \cr
& NGC~4051 & 0.52 & 0.24 & $-1.04$ $^S$ & 0.7&  0.5 &      &                                &  (IV) & \cr
& NGC~4123 & 0.36 & 0.30 & $-0.93$ $^S$ & 1.2&  0.8 &      &                                &  (IV) & \cr
& NGC~4303 & 0.63 & 0.10 & $-1.21$ $^S$ & 7.5&  5.3 & 14.0 & $-0.073$ $^{M_1}$              & II-III & \cr
& NGC~4304 & 0.54 & 0.34 & $-1.06$ $^D$ &    &      &      &                                &       & \cr
& NGC~4321 & 0.74 & 0.12 & $-1.33$ $^S$ & 5.7&  4.0 &      & $-0.035$ $^{M_1}$              &   II  & \cr
& NGC~5236 & 0.38 & 0.22 & $-1.03$ $^Y$ &13.4&  9.4 &      & $-0.024$ $^Z$                  &  III  & \cr
& NGC~5248 & 0.90 & 0.12 & $-1.22$ $^S$ & 5.3&  3.7 &  4.3 &                                &   II  & \cr
& NGC~5371 & 0.39 & 0.18 & $-1.25$ $^S$ & 5.6&  3.9 &      &                                & III-IV & \cr
& NGC~5457 & 0.86 & 0.05 & $-1.33$ $^S$ & 2.5&  1.8 &      & $-0.109$ $^{M_1}$              &    I  & \cr
& NGC~5921 & 0.34 & 0.25 & $-1.26$ $^D$ &    &      &      &                                &       & \cr
& NGC~6217 & 0.40 & 0.38 & $-1.03$ $^S$ & 3.2&  2.2 &  5.4 &                                & III-IV & \cr
& NGC~6384 & 0.64 & 0.12 & $-1.48$ $^Y$ & 4.0&  2.8 &      &                                &  (II) & \cr
& NGC~6744 & 0.52 & 0.12 & $-1.41$ $^R$ &    &      &      &                                &       & \cr
& NGC~6946 & 0.87 & 0.05 & $-1.24$ $^Y$ &10.0&  7.0 &  3.5 & $-0.089$ $^{M_1}$              &   II  & \cr
& NGC~7479 & 0.41 & 0.47 & $-0.80$ $^S$ &17.1& 12.0 &  8.1 &                                &  III  & \cr
& NGC~7678 & 0.47 & 0.23 & $-1.18$ $^S$ &10.9&  7.6 &      &                                & (III) & \cr
& NGC~7741 & 0.20 & 0.34 & $-1.39$ $^Y$ & 0.8&  0.6 &  1.8 &                                &   IV  & \cr
\noalign{\medskip\hrule\smallskip}
}}
\item{} Cols.~(2) and (3). From Martin (1995).
\item{} Col.~(4). From $^D$ $\rightarrow$ Devereux (1987); $^R$ $\rightarrow$
Rice et al. (1988); $^S$ $\rightarrow$ Soifer et al. (1989); $^Y$
$\rightarrow$ Young et al. (1989).
\item{} Col.~(5). From Eq.~2.
\item{} Col.~(6). From Eq.~1 using data by Young et al. (1989).
\item{} Col.~(7). From Kennicutt (1983).
\item{} Col.~(8). $^*$ $\rightarrow$ Respective slopes in the bar / in the
disc regions.  
From $^{M_1}$ $\rightarrow$ Martin \& Roy (1994); $^{M_2}$
$\rightarrow$ Martin \& Roy (1995); $^{R}$ $\rightarrow$ Roy
\& Walsh (1997); $^{W}$ $\rightarrow$ Walsh \& Roy (1996);
$^Z$ $\rightarrow$ Zaritsky et al. (1994).
\item{} Col.~(9). Classes indicated in brackets are uncertain.
\endtab

\titlea{The galaxy sample}
We have selected all the barred galaxies with 1) bar length $L$ and
bar axis ratio $(b/a)$ determined by Martin (1995), and 2) IRAS fluxes
from various authors. The basic source is Soifer et al. (1989) which
is completed from lists by Rice et al. (1988), Young et al. (1989), or
Devereux (1987).

Five early-type (Sa to Sb) and 42 late-type (Sbc to Scd) galaxies in
Martin's list have IRAS data. The sample was reduced, using two more
selection criteria: 1) Galaxies should not be too inclined so that the
deprojected axis ratio $(b/a)_i$ does not deviate too much from the
projected value (a difference of 1 ellipticity class as defined by
Martin was admitted).  2) Galaxies should be isolated or weakly
interacting (no companion detected within $10D_{25}$; no morphological
disturbances) in order to clearly separate effects coming from the bar
from those generated by interactions. An exception is NGC~5457 which
presents some moderate morphological disturbances.  Seyfert galaxies
have not been excluded from the sample, but only two are present
(NGC~1365 and NGC~4051).

We have 32 late-type and only 4 early-type objects left. Arguments
from the literature can incite to only consider the late types: Bars
in them could have spontaneously formed in discs contrary to bars in
early types, induced by interactions, as suggested by Noguchi
(1996). In fact, in Sect.~5, we will deal with spontaneous
bars. Moreover, Sauvage \& Thuan (1994) have found a decreasing
contribution of the cirrus component to $L_{\rm FIR}$ toward later
types. These issues are still subjects of debate.  Moreover, the
number of early-types that would otherwise be in our sample is small
(4) and their inclusion or exclusion does not affect the results and
the conclusions at all (see below), which will be based on a
final sample of 32 Sbc--Sc--Scd galaxies.

Available data on radial O/H gradients $d[{\rm O/H}]/dR$ have also
been collected from various sources but mainly from the tables of
Martin \& Roy (1994) and Zaritsky et al. (1994).

The selected galaxies cover nearly two decades in FIR luminosity ($1.1
\cdot 10^9$ to $8.9 \cdot 10^{10} \, \rm L_\odot$). NGC~3486 and
NGC~7479 are respectively the least and most luminous objects of the
sample.  The luminosity of NGC~7479 is similar for instance to M82 but
nearly two orders of magnitude below Arp~220.  Strong starbursts are
generally characterized by $L_{\rm FIR} \ga 10^{11} \, \rm L_\odot$
and result from significant interactions.  

Without being a strong starburst, NGC~7479 has photometric properties
comparable to average values given for ``Starburst Nucleus Galaxies''
(SBNG) as defined by Balzano (1983) or Coziol et al. (1994).  For
instance, $\log(S_{25}/S_{100}) \, \rm{(NGC~7479)} \!=\! -0.80$ very
close to the average value $-0.75$ inferred for SBNGs by
Dultzin-Hacyan et al. (1990). By comparison, we have for the three
following well-known starbursts:

\noindent
$\log(S_{25}/S_{100}) \, \rm{(NGC~253)} \!=\! -1.10$,

\noindent
$\log(S_{25}/S_{100}) \, \rm{(NGC~1614)} \!=\! -0.62$, 

\noindent
$\log(S_{25}/S_{100}) \, \rm{(M82)} \!=\! -0.68$.  

\noindent
Moreover, NGC~7479 has $f \!=\! 17.1$, whereas $f$ is respectively 6.6
and 143 for NGC~253 and NGC~1614.

The final sample of galaxies is listed in Table~1. The column entries
are as follows:

\noindent 
Col.~(1): NGC designation of the galaxy.

\noindent 
Col.~(2): Deprojected bar axis ratio.

\noindent 
Col.~(3): Deprojected bar length in $D_{25}$ unit.

\noindent 
Col.~(4): Logarithm of the $S_{25}/S_{100}$ IRAS color index.

\noindent 
Col.~(5): Relative star formation rate from FIR data (Eq.~2).

\noindent 
Col.~(6): Star formation rate from FIR data (Eq.~1) in 
[$\rm M_\odot \,yr^{-1}$].

\noindent 
Col.~(7): Star formation rate from H$\alpha$ emission in [$\rm M_\odot
\, yr^{-1}$].

\noindent 
Col.~(8): Radial O/H abundance gradient in [$\rm dex \, kpc^{-1}$].

\noindent 
Col.~(9): Class of the galaxy as defined in Sect.~6.

\titlea{Observed connections}
The following subsections discuss the connections or absence of
connections between various quantities referring either to the bar
morphology, or to the star formation activity, or to abundances
indices.  Some conventional terminologies are useful.  A bar with an
axis ratio $(b/a)_i \le 0.6$ will be called {\it strong} in the text,
whereas the term {\it weak} will be reserved to bars with $(b/a)_i
\!>\!  0.6$.  Similarly, a bar will be called {\it long} if its
relative length $2L_i/D_{25} \ge 0.18$ and {\it short} in the opposite
case.  Finally, we will consider two classes of FIR colours
($\log(S_{25}/S_{100} \ge -1.2$ or $\log(S_{25}/S_{100}) \!<\! -1.2$)
corresponding to more or less pronounced star formation activity.
Clearly, these chosen limits are somewhat arbitrary. However, slight
changes of these values do not affect the results presented below.
They have been chosen as follows: The axis ratio of 0.6 corresponds to
the middle of the interval of observed $(b/a)_i$ values, i.e 0.2 --
1.0. The length of 0.18 separates our sample in roughly half long bars
and half short bars. The FIR colour of $-1.2$ approximately separates
systems whose FIR emission is dominated either by star formation from
those dominated by cirrus (see Sect.~2.1.3).

\begfig 8.8 cm
\nompsy{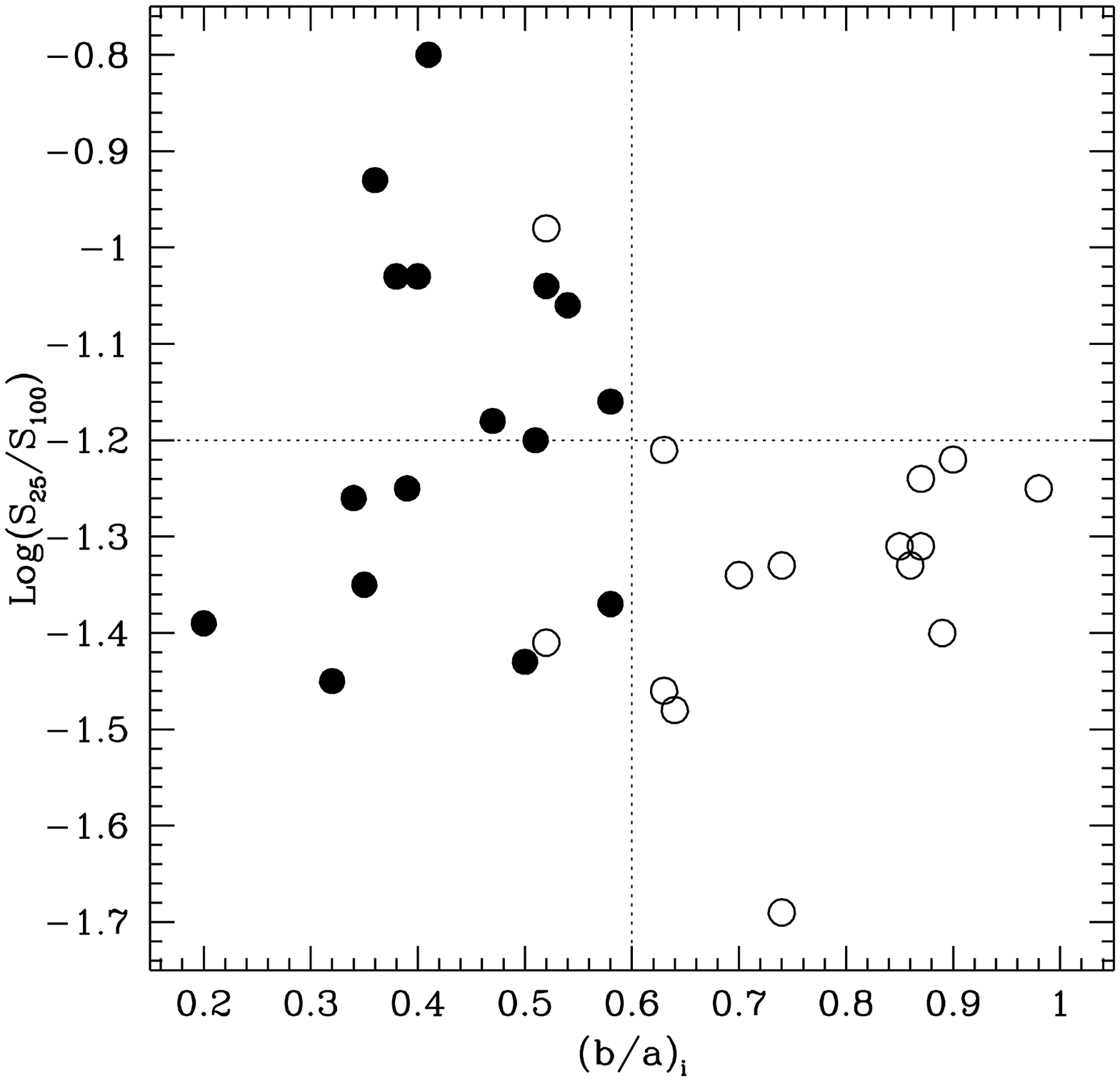}
\figure{\FBAS}{
Relation between the star formation indicator $\log(S_{25}/S_{100})$
and the deprojected bar axis ratio $(b/a)_i$ for the sample of
galaxies.  Full circles are for long bars ($2L_i/D_{25} \ge 0.18$) and
open circles for short bars ($2L_i/D_{25} \!<\! 0.18$).  Dotted lines
separate strong bars from weak ones as well as galaxies actively
forming stars from more quiescent ones }
\endfig


\titleb{Star formation activity -- bar strength}
The link between the $\log(S_{25}/S_{100})$ and the deprojected bar
axis ratio $(b/a)_i$ (the bar ``strength'' parameter) is shown in
Fig.~{\FBAS}.  In our sample, all the galaxies having
$\log(S_{25}/S_{100}) \ge -1.2$ have strong bars (10 galaxies). On the
contrary, all the weakly barred galaxies display low current star
formation activity (14 galaxies).  There are also strongly barred
galaxies which do not actively form stars (8 galaxies).  These
non-active strongly barred galaxies could be either in a
``pre-starburst'' or in a ``post-starburst'' phase (see Sect.~5).
Using other indicators of star formation activity mentioned in
Sect.~2.1.3 (e.g. $\log(S_{25}^2/S_{12}S_{100})$) does not alter the
tendency shown in Fig.~{\FBAS}.  With only 32 objects, our statistics
is still poor and one should remain cautious before drawing general
conclusions. However, in Fig.~{\FBAS} the different behaviour of weak
and strong bars is striking and consistent with the results of
numerical simulations presented in Sect.~5.  Moreover, Martin (1995)
had also noticed that the fraction of strong bars is higher in
galaxies with nuclear activity than in quiescent galaxies.

\begfig 8.8 cm
\nompsy{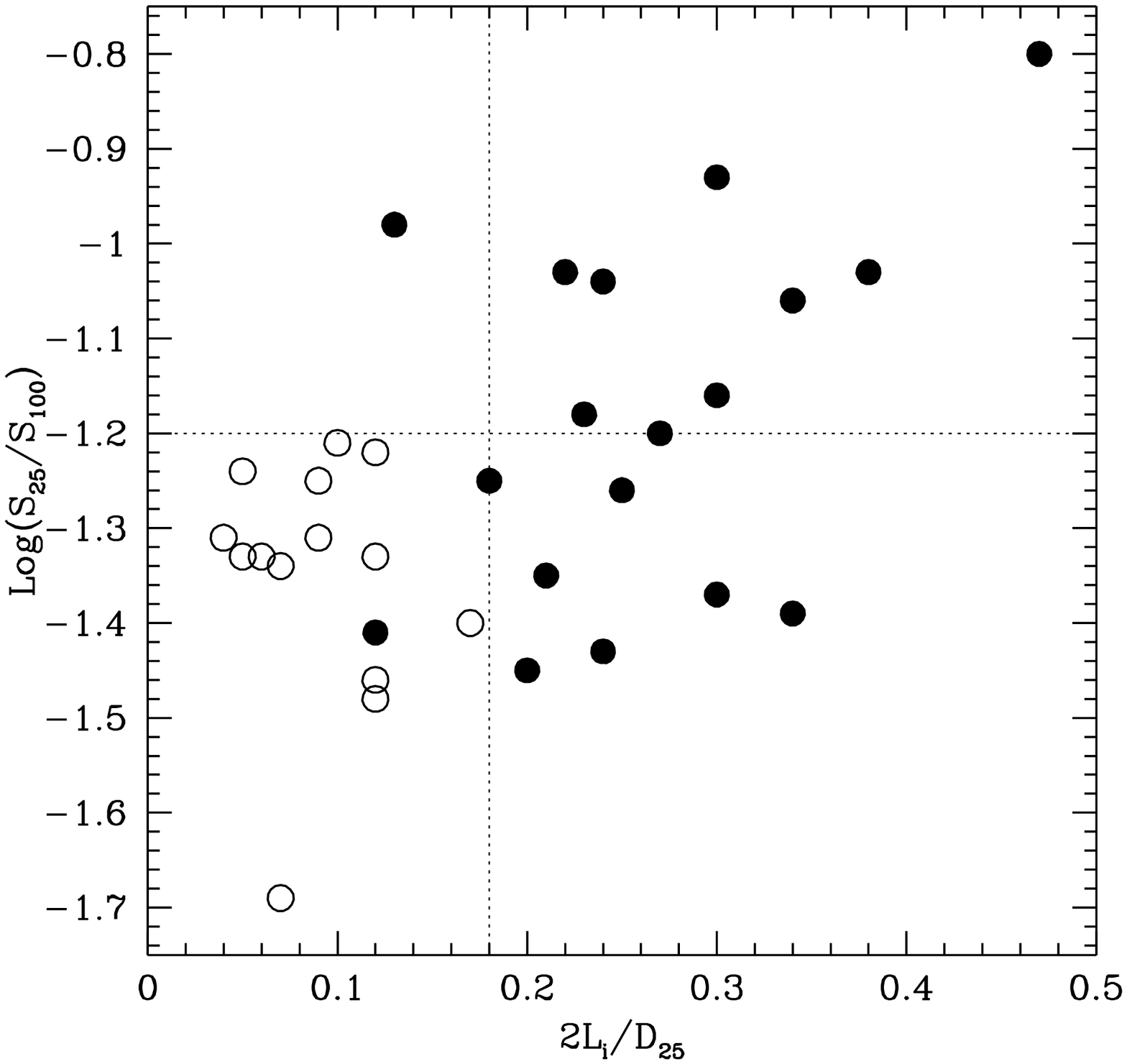}
\figure{\FLS}{
Relation between the $\log(S_{25}/S_{100})$ and the relative
deprojected bar length $2L_i/D_{25}$ for the sample of galaxies.  Full
circles are for strong bars ($(b/a)_i \le 0.60$) and open circles for
weak bars ($(b/a)_i \!>\! 0.60$).  Dotted lines separate long bars
from short ones as well as galaxies actively forming stars from more
quiescent ones }
\endfig

\begfig 8.8 cm
\nompsy{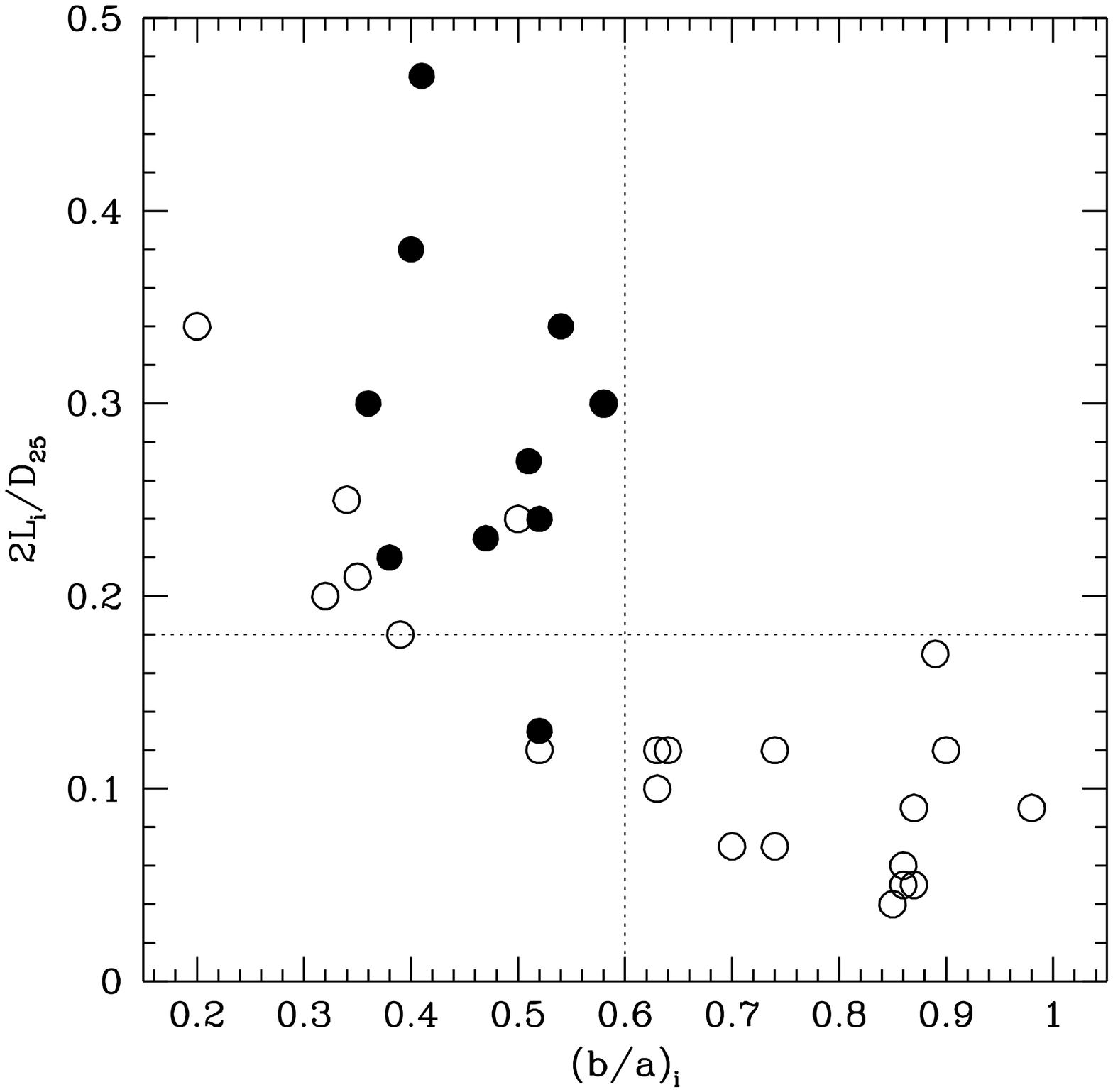}
\figure{\FBAL}{
Relation between $(b/a)_i$ and $2L_i/D_{25}$ for the sample of
galaxies.  Full circles are for galaxies actively forming stars
($\log(S_{25}/S_{100}) \ge -1.2$) and open circles for more quiescent
galaxies ($\log(S_{25}/S_{100}) \!<\! -1.2$).  Dotted lines separate
strong bars from weak ones as well as long bars from short ones }
\endfig

\titleb{Star formation activity -- bar length}
The link between the $\log(S_{25}/S_{100})$ and the relative
deprojected bar length $2L_i/D_{25}$ is shown in Fig.~{\FLS}.  All the
galaxies but one (NGC~1637) with $\log(S_{25}/S_{100}) \ge -1.2$ have
long bars (9 galaxies).  This can be explained as follows: Bar-driven
movement of gas towards the center takes place inside the co-rotation
radius, which is generally close to the end of the bar. Thus in those
systems with longer bars, a greater fraction of the total store of gas
in the system can be swept up and driven towards the center.
Similarly, all the galaxies with a short bar appear to be more
quiescent (15 galaxies).  However, there are also galaxies with low
star formation and long bars (7 galaxies).  Thus, the increase of the
bar length generally seems to have a similar effect as the decrease of
the bar axis ratio.

Surprisingly enough, these two quantities are strongly correlated in
our sample (see Fig.~{\FBAL}). Strong bars are long (except NGC~1637
and NGC~6744) and weak bars are short. Whereas it is well-known that
on average bars of early-type galaxies are longer than those of
late-type galaxies (e.g. Martin 1995), so far no correlation between
the length and strength of late-type bars seems to have been
highlighted.

\titleb{Abundance gradient -- bar strength}
Martin \& Roy (1994) established a correlation between the radial O/H
abundance gradient in the discs of SBs and the bar axis ratio $(b/a)$
in the sense that stronger bars have a rather flat gradient, whereas
steeper gradients are observed in galaxies with weak or no bars.
Taking into account the internal uncertainties and comparing with the
data by other authors (e.g. Vila-Costa \& Edmunds 1992; Zaritsky et
al. 1994), a rather larger dispersion of points in the diagram
$\delta_{\rm O} \equiv d[{\rm O/H}]/dR$ versus $(b/a)$ is observed but
the above general trend is clearly present.

This suggests that other parameters, such as the efficiency of star
formation $\epsilon \!=\!  L_{\rm FIR}/M_{\rm H_2}$ might play a role
in the connection between the chemical and dynamical evolution of
bars.  According to Tinsley (1980), in a quasi-stationary state the
radial chemical gradient essentially depends on the ratio of two
timescales, i.e.
$$
{d(Z/y) \over dR} \sim
{-R^{-1}} {\tau_{\rm in} \over \tau_{\rm sf}} \, ,
\eqno(\new)
$$
where $\tau_{\rm in}$ is the characteristic timescale of gas inflow
through the center, $\tau_{\rm sf}$ is the characteristic timescale
for exhausting gas through star formation, and $y$ is the yield.  The
extension of this model to the present context suggests that at first
approximation $\tau_{\rm in} \sim (b/a)$ since stronger bars have
higher gas mass inflow (as shown e.g. by Friedli \& Benz 1993).
Furthermore $\tau_{\rm sf} \sim \epsilon^{-1}$.  So, locally the
dependence of the chemical gradient on bar axis ratio must be weighted
by the star formation efficiency.

Due to the lack of data in our sample, we must restrict ourselves to
register a qualitative agreement between relative observed and
calculated gradients for NGC~3344, 4303, 4321, 5236, and 6946.  But
the real situation can even be more complicated. First, the simplified
formula above also shows an $R$ dependence. Second, in some galaxies
two different slopes for the O/H abundance gradient have recently
clearly been inferred, i.e. NGC~3359 (Martin \& Roy 1995) and NGC~1365
(Roy \& Walsh 1997). In these two galaxies, the abundance gradient is
flat in the disc region, whereas a moderate negative gradient subsists
in the bar region (see Table~1).  Note that very few galaxies have at
least 30 measured HII regions, a necessary condition to be in position
to highlight this feature.  The numerical simulations reported in
Sect.~5 show this feature and indicate that the age of the bar is
another factor influencing the chemical gradient.

\titleb{Connections with Hubble type}
The sample of Table~1 contains late-type galaxies with Hubble types
between T=3 and 7. No link has been found between the Hubble type and
either $(b/a)_i$, or $2L_i/D_{25}$, or $\log(S_{25}/S_{100})$.

\titlea{Clues from numerical simulations}
\titleb{Spontaneous bar formation in discs} 
The most recent ideas concerning the various processes which drive the
formation, the evolution and the destruction of bars can be found in
the thorough reviews by Sellwood \& Wilkinson (1993), Martinet (1995),
and Sellwood (1996).  Here, only the timescale problem for spontaneous
bar formation is briefly discussed.

The details of the formation of galactic discs are still under debate
(e.g. Dalcanton et al. 1997), but there are various evidences that
their formation could require a non-negligible fraction of Hubble
time, or could even be an ongoing process.  For instance, various
cosmological numerical simulations using the standard CDM scenario
indicate timescales of several Gyr for the disc formation
(e.g. Steinmetz \& M\"uller 1995). Based on chemical abundance
arguments, Sommer-Larsen \& Yoshii (1990) also found that continuous
infall of proto-galactic material onto the disc over timescales of
4--6~Gyr is necessary.  In addition, it is interesting to note that
late-type discs appear to be younger than early-type discs
(Sommer-Larsen 1996) which suggests that the beginning of the disc
assembly might not occur at an universal epoch.

The mechanism of bar formation requires the presence of a well defined
disc. The growth of a spontaneous bar may occur in later stages of the
disc evolution, i.e. when sufficient gas mass has been accreted onto
the disc and, above all, transformed into stars via star formation
processes.  At some point, the stellar disc will meet the critical
conditions necessary for the onset of the bar instability. Star
formation works towards this by progressively adding dynamically cool
masses in the stellar disc.  For instance, in the models of Noguchi
(1996), spontaneous bars typically appear only 6--7~Gyr after the
beginning of the disc formation.  This is about 10 times longer than
the growth of a strong bar (see Sect.~5.3).  Thus, the existence of
young bars among nearby galaxies is highly expected and very likely
observed (Martin \& Roy 1995; Martin \& Friedli 1997).  The fact that
barred galaxies seem to be scarce in the Hubble Deep Field (van den
Bergh et al. 1996) could also be considered as a possible confirmation
of the late appearance of such structures in the life of flattened
galaxies. However, this latter study only presents preliminary results
which clearly have to be confirmed and interpreted with great care.

\titleb{Method, previous and present models}
In order to try to explain the observed connection between the bar
strength and the SFR presented in the previous section, we have
performed a new set of 3D self-consistent numerical simulations with
stars, gas, and star formation.  Technical details can be found in
Pfenniger \& Friedli (1993), and Friedli \& Benz (1993, 1995). Since
our galaxy sample is made of non- or weakly-interacting galaxies, and
because the complete process of disc formation is clearly beyond the
scope of this paper, we here emphasize some steps of possible
evolution implying the mechanism of spontaneous bar formation in an
already existing disc.

Previous numerical models (Friedli \& Benz 1993) have indicated that
the intensity of the gas fueling phenomenon strongly depends on the
strength of the bar.  As the bar axis ratio $(b/a)$ decreases, much
faster gas accumulation into the center occurs. In less than one Gyr,
strong bars have nearly pushed all the gas initially inside the bar
region into the center, whereas weak bars have only accreted a small
fraction of it.  The formation of a spontaneous or induced strong bar
in a gas-rich Sc-like disc typically triggers a starburst of
intermediate power and duration.  Depending on the initial amount of
gas and the various parameters of the star formation ``recipe''
(Friedli \& Benz 1995), the peak star formation rate is $\approx 8-16
\, \rm M_\odot \, yr^{-1}$ and the duration is $\approx 0.1-0.2$ Gyr.
In particular, stars are formed in gaseous regions unstable with
respect to the Toomre parameter (Toomre 1964) $Q_g = s \kappa / \pi G
\Sigma_g$, where $s$ is the sound speed, $\kappa$ is the epicyclic
frequency, and $\Sigma_g$ is the gas surface density.  The star
formation parameters used in the numerical simulations are not unique
but have carefully been chosen in order to reproduce as accurately as
possible the observed star formation properties of barred galaxies.

Below, the computed SFR is the mean value over 100~Myr, i.e. from
50~Myr before to 50~Myr after the time at which the corresponding
$(b/a)_{\rm max}$ is determined.  This determination was done by
applying a standard ellipse-fitting routine to the stellar surface
density distribution.  The generic model has an initial Sc-like disc
with an initial gas to stars mass ratio $M_g/M_* \!=\! 0.11$, where
$M_g$ is the total gas mass and $M_*$ is the total stellar mass.  The
value of the Toomre parameter (Toomre 1964) $Q_* = \sigma_R \kappa /
3.36 G \Sigma_* \approx 1.7$, where $\sigma_R$ is the radial stellar
velocity dispersion, and $\Sigma_*$ is the stellar surface density.
The initial O/H abundance gradient has been set to $-0.1 \, \rm dex \,
kpc^{-1}$.  In order to smoothly switch on the star formation, the
models are first calculated during 400~Myr in a forced axisymmetric
state.  This procedure also allows us to compute at $t\!=\!0$ the SFRs
for the corresponding unbarred galaxy.  After that, the simulations
described below evolve in a fully self-consistent way.

\begfig 8.8 cm
\nompsy{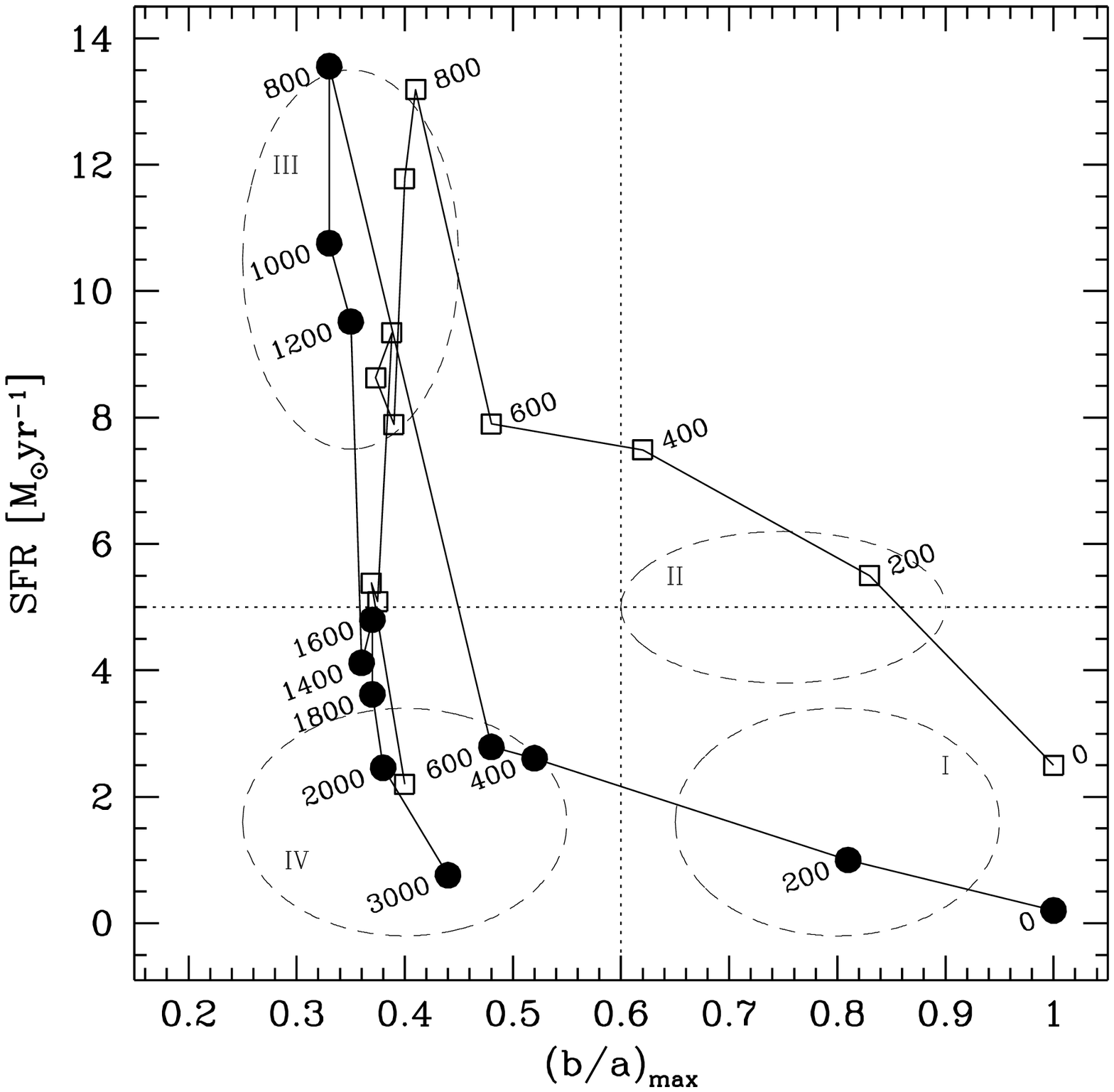}
\figure{\FBASFRT}{
Time evolution of the total SFR and the maximum bar axis ratio
$(b/a)_{\rm max}$ for typical numerical simulations forming a {\it
strong} bar, either with $M_g/M_* \!=\! 0.11$ (full circles; generic
model), or with $M_g/M_* \!=\! 0.17$ (open squares).  The time in Myr
is indicated beside symbols when possible.  Dotted lines separate
strong bars from weak ones as well as galaxies actively forming stars
from more quiescent ones. The dashed ellipses schematically indicate
the position of the four observational classes defined in Sect.~6 }
\endfig

\begfig 8.8 cm
\nompsy{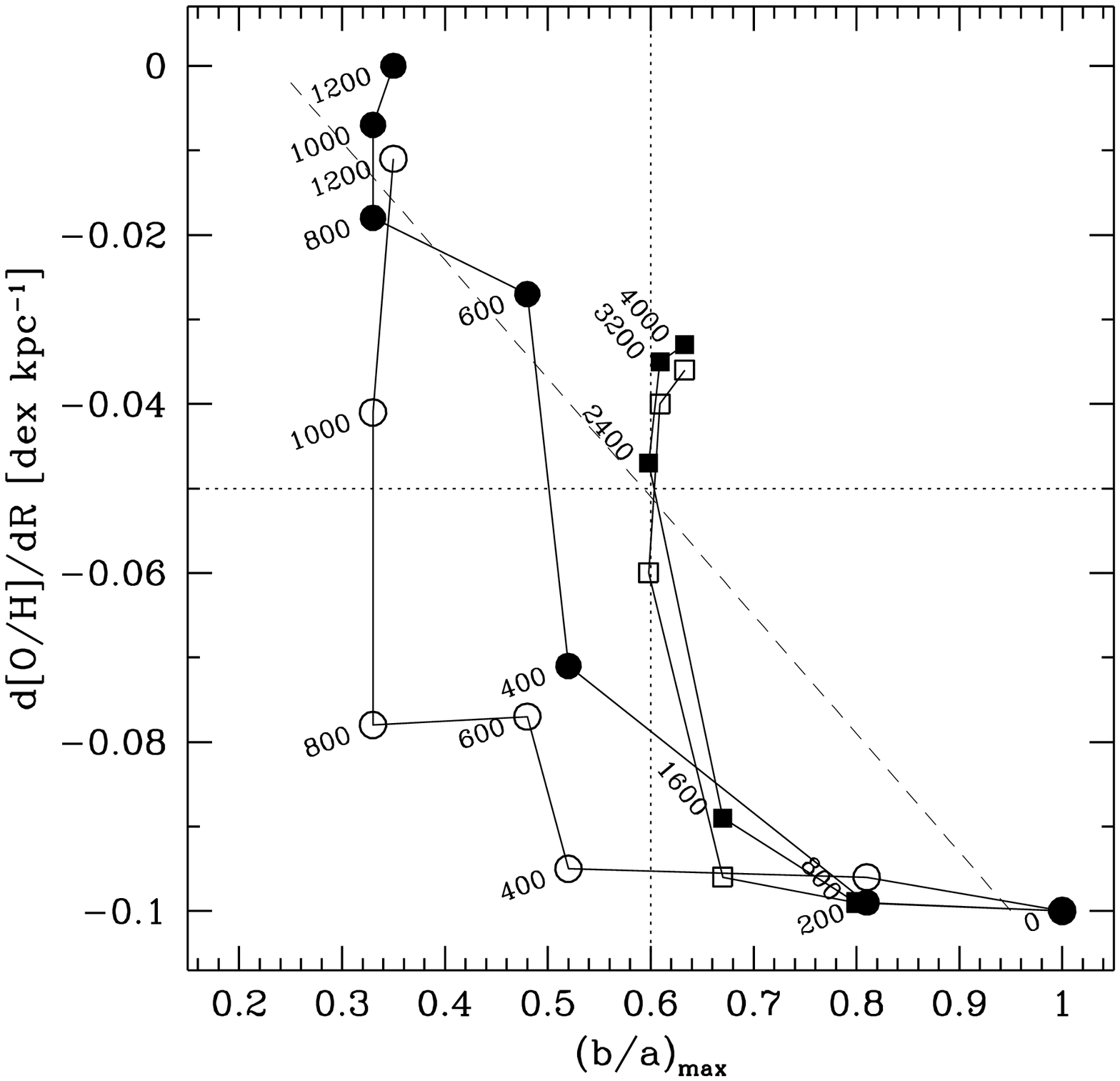}
\figure{\FBAZT}{
Time evolution of the O/H abundance gradient $\delta_{\rm O}$ and the
maximum bar axis ratio $(b/a)_{\rm max}$.  Circles corresponds to the
strong bar case (generic model) whereas squares are for the weak bar
case.  Open and full symbols are respectively for the abundance
gradients in the bar (i.e. 2 -- 8~kpc) and disc (i.e. 12 -- 18~kpc)
regions.  Both models have $M_g/M_* \!=\! 0.11$.  The time in Myr is
indicated beside symbols when possible.  Dotted lines separate strong
bars from weak ones as well as steep abundance gradients from shallow
ones. The dashed line corresponds to the observed relation by Martin
\& Roy (1994) }
\endfig

\titleb{Strong bar case}
For two representative simulations, Fig.~{\FBASFRT} shows the time
evolution of both the total SFR and the maximum bar axis ratio
$(b/a)_{\rm max}$.  As time is evolving, a bar instability
progressively develops triggering more star formation.  Note that the
bar growth and evolution timescales quoted below can be either shorter
or longer depending on the instability level of the stellar disc at
the beginning of the simulation.

For the generic model at times $t\!=\!400$~Myr and $t\!=\!600$~Myr,
the bar is already strong but the SFR is still modest, mainly
concentrated along the bar major axis.  This is the ``pre-starburst''
phase where not enough gas mass has been pushed into the center to
exceed the critical gas surface density $\Sigma_c$ necessary for the
onset of star formation.  The bar axis ratio is progressively
decreasing, whereas the bar length is gradually increasing up to $t
\approx 800$~Myr where the bar reaches a quasi-stable state
($(b/a)_{\rm max} \approx 0.33$).  This is the lowest bar axis ratio
of the whole simulation and it coincides with the maximum SFR observed
($13.6 \, \rm M_\odot \, yr^{-1}$) with the star formation essentially
concentrated at the center.  By $t \approx 1400$~Myr the SFR has
become more moderate once more, although the bar is still strong. This
is the beginning of the ``post-starburst'' phase where the gas has
been sufficiently consumed to go again below $\Sigma_c$ and nearly
stop star formation although large amounts of gas remain near the
center.

By $t\!=\!2000$~Myr, the bar axis ratio has increased very slightly
($(b/a)_{\rm max} \approx 0.38$), but the SFR has dropped by a factor
of more than 6 ($\rm SFR \approx 2.5 \, \rm M_\odot \, yr^{-1}$).
Finally, after one further Gyr, at $t\!=\!3000$~Myr, the SFR has
decreased to less than $1 \, \rm M_\odot \, yr^{-1}$, while the bar
has become a little weaker still ($(b/a)_{\rm max} \approx 0.44$).  At
this time, the total gas to star mass ratio $M_g/M_* \approx 0.04$,
and the gas represents less than one percent of the dynamical mass
inside 1~kpc.

So, clearly strong bars do not necessarily always host enhanced
central star formation.  The presence of observed galaxies in the
lower left corner of Fig.~{\FBAS} is thus easily explained.  The upper
right corner of Fig.~{\FBAS} cannot be reached and crossed by the
generic model.  However, one possible way is to strongly increase the
gas mass ($M_g/M_* \ga 0.15$) in order to produce a widely
over-critical disc of gas with respect to the Toomre criterion. Such
discs will form stars at a very high rate whatever the bar strength
is, and the formation of a strong bar only results in a moderate
increase of the total SFR.  As an example, the evolutionary track of a
model similar to the generic one but with $M_g/M_* \!=\! 0.17$ is also
presented in Fig.~{\FBASFRT}.

For the generic simulation, the time evolution of the radial O/H
abundance gradient $\delta_{\rm O}$ and the maximum bar axis ratio
$(b/a)_{\rm max}$ is presented in Fig.~{\FBAZT}.  The time evolution
of $\delta_{\rm O}$ in the bar (i.e. 2 -- 8~kpc) and disc (i.e. 12 --
18~kpc) regions are both shown. A very different behaviour is
observed. The disc abundance gradient becomes very quickly very
shallow as soon as the strong bar develops. It moves from $-0.10$ at
$t\!=\!200$~Myr to $\approx -0.02$ at $t\!=\!800$~Myr.  On the
contrary, the abundance gradient in the bar region remains first
steep.  It changes from $-0.10$ at $t\!=\!200$~Myr to $\approx -0.08$
at $t\!=\!800$~Myr, and only becomes shallower ($-0.01$) around
$t\!=\!1200$~Myr. Moreover, the galaxy core becomes very oxygen-rich
as well.

In the bar region, during the early phase of its existence, the
gradient is maintained since the gas dilution (following the
significant gas inflow) is compensating for the heavy-element
production in the furious star formation then going on in the nuclear
vicinity.  For more details, see also Friedli et al. (1994), Friedli
\& Benz (1995), Martin \& Friedli (1997).  This of course results in
the presence of {\sl two different radial abundance gradients in young
strongly barred galaxies}.  For instance, at $t\!=\!800$~Myr, the
slope ratio is about 4.3. After $t\!=\!1200$, the disc abundance
gradient remains essentially flat, whereas the lack of gas inside the
bar region prevents there any reliable determination of the abundance
gradient.  So, there is a ``steep-shallow'' break in the slope profile
of $\delta_{\rm O}$ as already observed in at least two galaxies (see
Table~1).  A ``shallow-steep'' break could also be present close to
the edge of the optical disc (see e.g. Friedli et al. 1994; Roy \&
Walsh 1997) but it has not yet been observed.

\begfig 8.8 cm
\nompsy{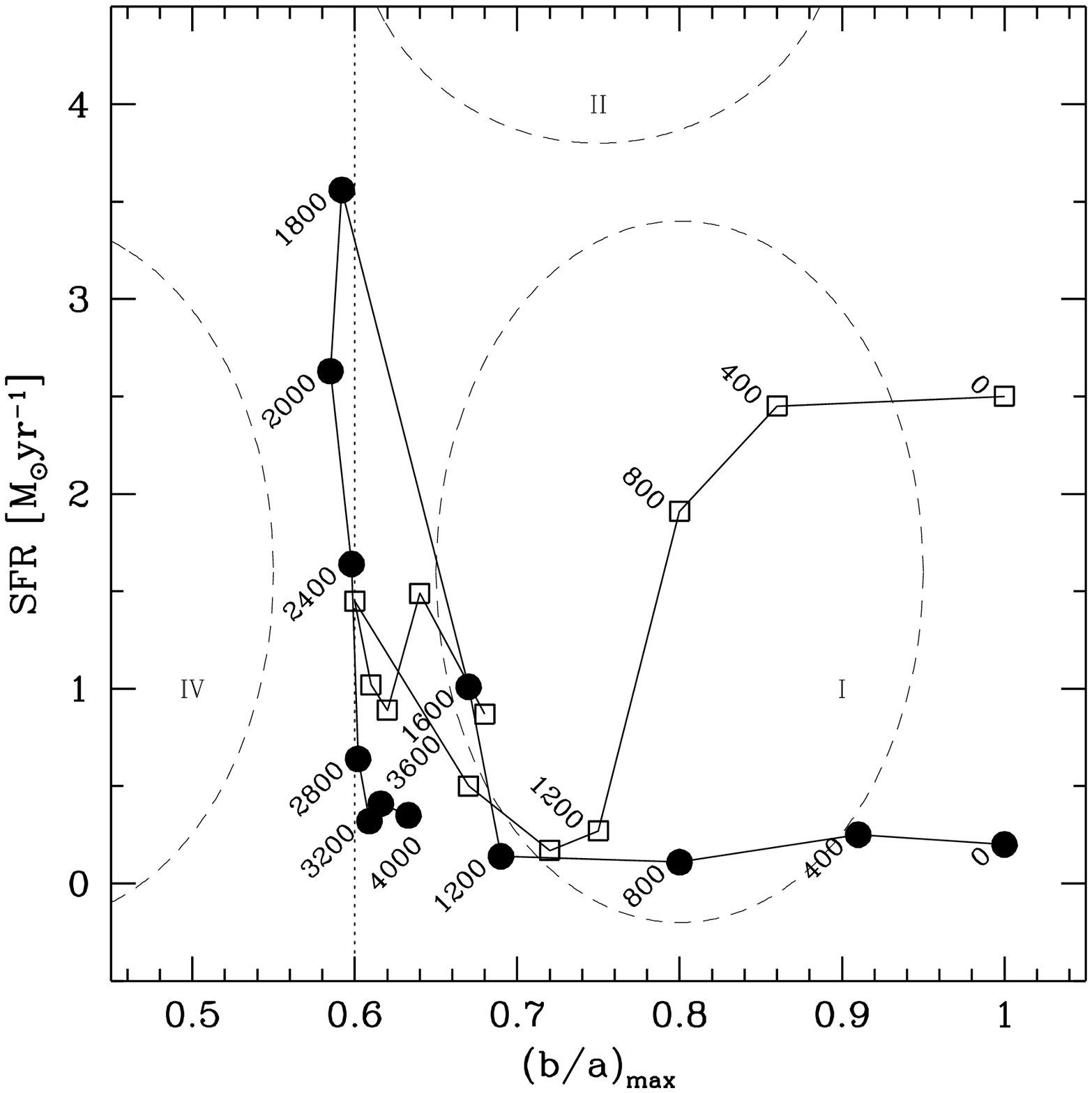}
\figure{\FBASFRTW}{
The same as for Fig.~{\FBASFRT} but for typical numerical simulations
forming a {\it weak} bar, either with $M_g/M_* \!=\! 0.11$ (full
circles), or with $M_g/M_* \!=\! 0.17$ (open squares). Note the
different scales for both axes with respect to Fig.~{\FBASFRT} }
\endfig

\titleb{Weak bar case}
Observed weak bars are either i) asymptotically and intrinsically
weak, or ii) transient features progressively turned into strong bars
as seen in the previous section. In the latter case, the duration of
the weak bar phase depends on the timescale of the bar instability
growth.  This timescale becomes longer if more of the total mass
resides in slow or non-rotating components like massive dark halos or
stellar bulges, i.e. the lower values of the Ostriker-Peebles
parameter (Ostriker \& Peebles 1973).  The fact that weak bars are
short (see Sect.~4.2) is an indication that they might in fact be
growing strong bars.  Moreover, it has been proven quite difficult to
numerically form permanent, realistic, weak bars.  They can however be
produced either by putting at the center a ``point mass'', e.g. a
dense cluster or a supermassive black hole with mass $0.01 \la M_{\rm
BH}/M_* \la 0.03$ (Friedli 1994; see also Norman et al. 1996), or by
increasing the stellar radial velocity dispersion so that $2 \la Q_*
\la 3$ (Athanassoula 1983).  The presence of large amounts of {\it
highly viscous} gas significantly reduces the maximum bar strength as
well (see Figs.~{\FBASFRT} and ~{\FBASFRTW}).

We chose the method of disc heating. So, the models presented here are
similar to the ones of Sect.~5.3 except for the $Q_*$ value which has
been increased by 30\%, i.e. $Q_* \approx 2.1$.  The time evolution of
both the total SFR and $(b/a)_{\rm max}$ is shown in
Fig.~{\FBASFRTW}. The bar growth timescale is much smaller than for
the strong bar case. The spiral arms also remain very weak since the
transfer of angular momentum is modest.

For the model with $M_g/M_* \!=\! 0.11$, up to $t\!=\!1200$~Myr, the
SFR is essentially constant and very low ($\approx 0.2 \, \rm M_\odot
\, yr^{-1}$), whereas the bar length is gradually increasing
and the bar axis ratio is progressively decreasing ($(b/a)_{\rm max}
\approx 0.69$).  Then, up to $t\!=\!2400$~Myr, the SFR appreciably
increases and the bar becomes a little stronger.  The maximum SFR is
$\approx 3.6 \, \rm M_\odot \, yr^{-1}$, the star formation being
essentially concentrated along the bar major axis and in the center.
The lowest bar axis ratio of the whole simulation is $(b/a)_{\rm max}
\approx 0.59$ so that this model corresponds in fact to an
intermediate case between weak and strong bars.  After that, up to
$t\!=\!4000$~Myr, the SFR continues to decrease to reach a
quasi-stationary state of $\approx 0.4 \, \rm M_\odot \, yr^{-1}$
whose major contribution comes from a nuclear ring.  The bar axis
ratio increases up to $(b/a)_{\rm max} \approx 0.63$.  At the end of
the simulation, $M_g/M_* \approx 0.09$.

The evolution of the model with $M_g/M_* \!=\! 0.17$ is somewhat
surprising.  Up to $t\!=\!800$~Myr, the over-critical gaseous disc
forms stars at a relatively high rate ($\approx 2.5 \, \rm M_\odot \,
yr^{-1}$). Then, the SFR suddenly drops as the disc is progressively
becoming self-regulated, i.e. $Q_g \ga 1$ all over the gaseous disc.
At $t \!=\! 2000$~Myr, the growth of the weak bar (up to $(b/a)_{\rm
max} \approx 0.60$) starts to influence the SFR which
increases. However, although high amounts of gas are still present,
the peak SFR of this model is smaller by a factor 2.5 than the one of
the model with $M_g/M_* \!=\! 0.11$.  The efficiency of gas fueling is
much reduced since the bar remains short, and its maximum axis ratio
gradually increases ($(b/a)_{\rm max} \approx 0.68$ at $t \!=\!
4000$~Myr). Thus, in our models, the increase of the $M_g/M_*$ ratio
results in shorter and weaker bars.

The time evolution of $\delta_{\rm O}$ and $(b/a)_{\rm max}$ is
presented for both bar and disc regions in Fig.~{\FBAZT}.  Contrary to
the strong bar case, a similar behaviour is observed in these two
regions although the abundance gradient remains always a little
steeper in the bar (at most by 27\%).  The abundance gradients
essentially remain unchanged up to $t \approx 1600$~Myr where
$(b/a)_{\rm max} \approx 0.67$.  Then, when $(b/a)_{\rm max} \la 0.6$,
they relatively quickly flatten to finally reach a nearly constant
value around $-0.035$ at $t \!=\! 4000$~Myr.  This value is slightly
larger than the one derived from the observed relation given by Martin
\& Roy (1994), i.e. $\delta_{\rm O} \approx -0.05$.  The relevant
points are that, a) weak bars need much longer timescales to flatten
radial abundance gradients, b) weak bars are unable to produce totally
flat abundance gradients.

\titlea{Discussion}
In order to quantitatively compare the results of the numerical
simulations with the observational features, we have tentatively
distinguished in our sample four main classes of late-type objects
according to their $(b/a)_i$, SFR/relative SFR ($f$), and
$\log(S_{25}/S_{100})$. We stress that only relative values of the
last three parameters are reliable.  The respective properties of the
four classes have been summarized in Table~2.  Clearly the galaxies
belonging to these various classes occupy different regions of this 4D
parameter space.  As can be seen in the suggested classification of
Table~1, some galaxies have either intermediate properties between two
classes, or half the properties of one class and the other half from
another class.  Of course, some galaxies cannot be classified due to
the lack of data.  For the same reason, we decided not to include
the abundance gradient as a fifth parameter in this classification.

\begtabfull
\tabcap{2}{Characteristics of galaxy class}
\vbox{
\halign{
& # & #\hfill \quad & \hfill#\hfill \quad & \hfill#\hfill \quad &
\hfill#\hfill \quad & \hfill#\hfill \quad & # \cr
\noalign{\hrule\medskip}
& \hfill Parameter \hfill
& \hfill Class I \hfill 
& \hfill Class II \hfill 
& \hfill Class III \hfill 
& \hfill Class IV \hfill &\cr
\noalign{\medskip\hrule\medskip}
& Bar strength           & weak          & weak           & strong      & strong     & \cr
& SFR                    & low           & median         & high        & low        & \cr
& $f$                    & $\la 3$       & $4-10$         & $\ga 10$    & $\la 2$    & \cr
& $\log(S_{25}/S_{100})$ & $\la -1.3$    & $\approx -1.3$ & $\ga -1.0$  & $\la -1.3$ & \cr
\noalign{\medskip\hrule\smallskip}
}}
\endtab

\noindent
{\it Class~I.} Galaxies with large $(b/a)_i$ (0.65 -- 0.95), weak SFR
($\la 3 \, \rm M_\odot \, yr^{-1}$), $f \la 3$, and very low values of
$\log(S_{25}/S_{100})$ ($\la -1.3$). Typical galaxies of this class
are NGC~3344, NGC~3726 and NGC~5457.

\noindent
{\it Class~II.} Galaxies are here again characterized by large
$(b/a)_i$ (0.60 -- 0.90), but much higher SFR ($\approx 4-6 \, \rm
M_\odot \, yr^{-1}$), and $f$ (4 -- 10).  The $\log(S_{25}/S_{100})$
is between $-1.2$ and $-1.3$. Representative galaxies of this class
are NGC~4321, NGC~5248, and NGC~6946.

\noindent
{\it Class~III.} This class includes galaxies with strong bars
$(b/a)_i$ (0.25 -- 0.45), very high SFR ($\ga 6 \, \rm M_\odot \,
yr^{-1}$), $f \ga 10$, and $\log(S_{25}/S_{100}) \ga -1.0$. Two
typical objects are NGC~5236 and NGC~7479.

\noindent
{\it Class~IV.} As in the previous class, galaxies belonging to this
last class have strong bars $(b/a)_i$ (0.25 -- 0.55), but much weaker
SFR ($\la 3 \, \rm M_\odot \, yr^{-1}$), $f \la 2$, and
$\log(S_{25}/S_{100}) \la -1.3$. Prototypes are NGC~3359, NGC~3887 and
NGC~7741.

The four SBa--SBb objects initially selected (see Sect.~3) have a
strong bar. The value of $\log S_{25}/S_{100}$ for NGC~3351 ($(b/a)_i
\!=\! 0.56$) is $-1.13$. For NGC~7552 ($(b/a)_i \!=\! 0.29$), it is
$-0.92$. Both could be included in class~III. NGC~4394 ($(b/a)_i \!=\!
0.48$) and NGC~4725 ($(b/a)_i \!=\! 0.43$) respectively have $-1.35$
and $-1.42$. They should belong to class~IV. As already noted, they do
not modify the discussion in this paper.

Of course, the transition from one class to the other is not
instantaneous and some galaxies can populate regions outside the
schematic and indicative zones drawn on Figs.~{\FBASFRT} and
{\FBASFRTW}.  The confrontation with the results from numerical
simulations of galaxies developing spontaneous bars clearly suggests
possible evolutionary sequences as presented below.

\noindent
{\it i) Strong bar:} When $M_g/M_* \approx 0.1$, the sequence should
be [I $\rightarrow$ IV $\rightarrow$ III $\rightarrow$ IV].  In the
case where the gas to star mass ratio is higher ($M_g/M_* \ga 0.15$),
the evolutionary sequence is [II $\rightarrow$ III $\rightarrow$ IV
\fonote{or even I if central gas accumulation nearly dissolves the bar
(see e.g. Friedli \& Benz 1993).}] instead.  A subsequent evolution
[IV $\rightarrow$ III] could only be considered if high amounts of
fresh gas are provided to the bar over short timescales, i.e. most
likely from outside.

\noindent
{\it ii) Weak bar:} The evolution is essentially enclosed in zone [I]
irrespective of the gas to star mass ratio. However, when $M_g/M_*
\approx 0.1$, a mini-starburst allows the model to come very close to
zone [II].

Class~IV is hybrid. Indeed, galaxies belonging to this class are
either in the ``pre-starburst'' phase or in the ``post-starburst''
one. The respective fraction of either type for the late-type galaxies
is unknown. However, the ``post-starburst'' galaxies should certainly
be more numerous since the ``pre-starburst'' phase appears quite short
in numerical simulations. There are at least two observational
possibilities to distinguish between these two phases. The first one
relates to star formation and the second one to abundance gradient.
In ``pre-starburst'' galaxies, star formation is increasing so that
current star formation should already be higher than recent star
formation. For instance $L_{\rm FIR}/L_{\rm B}$ should be high (Tomita
et al. 1996). According to numerical models (Sect.~5), these galaxies
show widely different abundance gradients in the bar ($\delta_{\rm O}$
$\rightarrow$ steep) and disc ($\delta_{\rm O}$ $\rightarrow$ shallow)
regions.  On the contrary, in ``post-starburst'' galaxies, star
formation is strongly declining so that current star formation should
already be lower than recent star formation, i.e. $L_{\rm FIR}/L_{\rm
B}$ should be low. These galaxies present similar abundance gradients
in the bar and disc regions ($\delta_{\rm O}$ $\rightarrow$ flat).
For instance in our sample, NGC~3359 has two different radial
abundance slopes; this galaxy seems to be in a ``pre-starburst'' phase
and its bar should be young as already suggested by Martin \& Roy
(1995).

The formation of a spontaneous or induced strong bar in a gas-rich
Sc-like disc appears to be a major dynamical event. It results in many
secular alterations of the galaxy properties over typical timescales
of one tenth of the Hubble time.  In this paper, we have mainly
focused on bar-induced changes of the star formation (both in spatial
distribution and intensity) and the radial gaseous abundance gradient.
However, bars are clearly able to deeply reshape the overall
morphology, kinematics, and chemistry of disc galaxies on less than a
Hubble time as well (see e.g. Martinet 1995).

\titlea{Conclusions}
The main aim of this paper was to improve our understanding of the
connection between star formation activity and the presence of bars in
spiral galaxies located in regions of low galaxy density.  In
particular, we intended to explain for what reasons only a fraction of
SBs show indices of enhanced star formation activity.  Our main
results can be summarized as follows:

\noindent
1) The non-interacting late-type galaxies most active in forming stars
have both strong $(b/a \le 0.6)$ and long $(2L_i/D_{25} \ge 0.18)$
bars. However, not all strong and long bars are actively creating
stars.  Weak bars do not display any significant excess of star
formation activity.  In general strong bars are long as well.
Although the sample is too poor to draw definite and more detailed
conclusions, the trend outlined here is unquestionable.

\noindent
2) The selected galaxies have been shared in four distinct classes
according to their respective bar strength, $\log(S_{25}/S_{100})$,
and relative SFR.  During the dynamical evolution, SBs probably go
through these classes according to various specific tracks.  Numerical
simulations enlight  possible scenarios. Very young strong bars are
first characterized by a vigorous episode of star formation and two
different radial gaseous abundance gradients, one steep in the bar and
one shallow in the disc.  Then, the galaxies progressively fall back
in a more quiescent state with a nearly flat abundance gradient across
the whole galaxy.  On the contrary, weak bars are unable to induce
significant star formation or flat abundance gradients.

\noindent
3) The slope profile of the radial abundance gradient is monitored by:
i) The strength of the bar (weak bar $\rightarrow$ slow and moderate
modifications of the initial gradient; strong bar $\rightarrow$ quick
flattening of the initial gradient).  ii) The age of the bar (young
bar $\rightarrow$ spatially distinct gradients; old bar $\rightarrow$
single gradient).  iii) The spatially-dependent star formation
efficiency.

\noindent
4) For late-type spirals the controversy concerning the role of bars
in enhancing/reducing star formation may be resolved if in fact only
{\it young and strong bars enhance star formation}. Larger samples
would however be necessary to fully confirm this assertion.

From these conclusions, it clearly appears that the formation of a
spontaneous  strong bar in an isolated  gas-rich Sc-like disc is a
major event in the dynamical history of the galaxy.  This results in a
significant and specific alteration of the spatial distribution and
intensity of star formation, as well as of the radial abundance
gradients.  In particular, some of these remarkable characteristics
could help finding young bars.

Finally, some caveats must be mentioned: a) These results only concern
isolated late-type spirals. b) We do not take into account possible
accretions or interactions which could play a role in the subsequent
internal evolution and consequently modify the star formation
efficiency, the morphology of bars, and even the Hubble type
(Pfenniger 1993).  c) The IRAS data do not have a sufficient
resolution to allow detailed studies of the star formation morphology,
e.g. along the bar or in the nucleus. Generally, the most intense star
formation tends to occur in the circumnuclear regions near the ILRs
when they exist or in the nuclei when ILRs are absent (see Telesco et
al. 1993). Studies of individual galaxies are still too scarce to
allow statistical investigations in such a context. No doubt that ISO
data will be most valuable to improve the present situation.

\acknow{
This work has been supported by the University of Geneva (Geneva
Observatory) and the Swiss National Science Foundation (FNRS).  We
especially thank D.~Pfenniger and J-R.~Roy for instructive discussions
and a careful reading of the manuscript as well as the referee,
T.G.~Hawarden, for very useful and detailed comments. }

\begref{References}

\ref
Arsenault R., 1989, A\&A 217, 66

\ref
Athanassoula, L., 1983, in: Internal Kinematics and Dynamics of
Galaxies, IAU Symp.~No.~100, ed.~L.~Athanassoula. Reidel, Dordrecht,
p.~243

\ref
Balzano V.A., 1983, ApJ 268, 602

\ref
Becklin E.E., 1986, in: Light on Dark Matter, ed.~F.P.~Isra\"el.
Reidel, Dordrecht, p.~415

\ref
Courteau S., de Jong R.S., Broeils A.H., 1996, ApJ 457, L73

\ref
Coziol R., Demers S., Peña M., Barneoud R., 1994, AJ 108, 405

\ref
Dalcanton J.J, Spergel D.N., Summers F.J., 1997, ApJ, in press

\ref
Devereux N., 1987, ApJ 323, 91 

\ref
Devereux N., Young J.S., 1990, ApJ 350, L25

\ref
Donas J., Deharveng J.M., Milliard B., Laget M., Huguenin D., 1987,
A\&A 180, 12

\ref
Dressel L.L., 1988, ApJ 329, L69

\ref
Dultzin-Hacyan D., Masegosa J., Moles M., 1990, A\&A 238, 28

\ref
Eskridge P.B., Pogge R.W., 1991, AJ 101, 2056

\ref
Friedli, D., 1994, in: Mass-Transfer Induced Activity in Galaxies,
ed.~I.~Shlosman. Cambridge University Press, Cambridge, p.~268

\ref
Friedli D., Benz W., 1993, A\&A 268, 65

\ref
Friedli D., Benz W., 1995, A\&A 301, 649

\ref
Friedli D., Benz W., Kennicutt R., 1994, ApJ 430, L105

\ref
Friedli D., Martinet L., 1996, in: Starburst Activity in Galaxies,
eds.~J.~Franco et al. RevMexAA (Serie de Conferencias), in press

\ref
Hawarden T.G., Mountain C.M., Leggett S.K., Puxley P.J., 1986, MNRAS
221, 41p

\ref
Hawarden T.G., Huang J.H., Gu Q.S., 1996, in: Barred Galaxies,
eds.~R.~Buta et al. ASP Conf. Series Vol.~91, p.~54

\ref
Huang J.H., Gu Q.S., Su H.J., et al., 1996, A\&A 313, 13

\ref
Helou G., 1991, in: The Interpretation of Modern Synthesis
Observations of Spiral Galaxies, eds.~N.~Duric, P.C.~Crane. ASP
Conf. Series Vol.~18, p.~125

\ref
Isobe T., Feigelson E.D., 1992, ApJS 79, 197

\ref
Keel W.C., 1983, ApJS 52, 229

\ref
Kennicutt R.C., 1983, ApJ 272, 54

\ref
Kennicutt R.C., 1989, ApJ 344, 685

\ref
Kennicutt R.C., Kent S.M., 1983, AJ 88, 1094

\ref
Kormendy J.J., 1982, in: Morphology and Dynamics of Galaxies, 12th
Advanced Course of the SSAA, eds.~L.~Martinet, M.~Mayor. Geneva
Observatory, Geneva, p.~113

\ref
Lonsdale Persson C.J., Helou G., 1987, ApJ 314, 513

\ref
Martin P., 1995, AJ 109, 2428 

\ref
Martin P., Friedli D., 1997, A\&A, in preparation

\ref
Martin P., Roy J-R., 1994, ApJ 424, 599 

\ref
Martin P., Roy J-R., 1995, ApJ 445, 161 

\ref
Martinet L., 1995, Fund. Cosmic Physics 15, 341

\ref
Moles M., M\'arquez I., P\'erez E., 1995, ApJ 438, 604

\ref
Noguchi M., 1996, ApJ 469, 605

\ref
Norman C.A., Sellwood J.A., Hasan H., 1996, ApJ 462, 114

\ref
Ostriker J.P., Peebles P.J.E., 1973, ApJ 186, 467

\ref
Pfenniger D., 1993, in: Galactic Bulges, IAU Symp.~No.~153,
eds.~H.~Dejonghe, H.J.~Habing. Reidel, Dordrecht, p.~387

\ref
Pfenniger D., 1996, in: Barred Galaxies and Circumnuclear Activity,
Nobel Symp.~98, eds.~Aa.~Sandqvist, P.O.~Lindblad. Lecture Notes in
Physics, Vol.~474, Springer, p.~91

\ref
Pfenniger D., Friedli D., 1993, A\&A 270, 561

\ref 
Pfenniger, D., Norman, C., 1990, ApJ 363, 391

\ref
Pogge R.W., 1989, ApJS 71, 433

\ref
Pompea S.M., Rieke G.H., 1990, ApJ 356, 416

\ref
Puxley P.J., Hawarden T.G., Mountain C.M., 1988, MNRAS 231, 465

\ref
Rice W., Lonsdale C.J., Soifer B.T., et al., 1988, ApJS 68, 91

\ref
Rice W., Boulanger F., Viallefond F., et al., 1990, ApJ 358, 418

\ref
Rowan-Robinson M., Crawford J., 1989, MNRAS 238, 523

\ref
Roy J-R., Walsh J.R., 1997, MNRAS, submitted

\ref
Sauvage M., Thuan T.X., 1992, ApJ 396, L69

\ref
Sauvage M., Thuan T.X., 1994, ApJ 429, 153

\ref
Sekiguchi K., 1987, ApJ 316, 145

\ref 
Sellwood J., 1996, in: Barred Galaxies, IAU Coll.~No.~157,
eds.~R.~Buta et al. ASP Conference Series, p.~259

\ref
Sellwood J., Wilkinson A., 1993, Rep. Prog. Phys. 56, 173

\ref
Soifer B.T., Boehmer L., Neugebauer G., Sanders D.B., 1989, AJ 68, 766

\ref
Sommer-Larsen J., 1996, ApJ 457, 118

\ref
Sommer-Larsen J., Yoshii Y., 1990, MNRAS 243, 468

\ref
Steinmetz M., M\"uller E., 1995, MNRAS 276, 549

\ref
Telesco C.M., 1988, ARA\&A 26, 343

\ref
Telesco C.M., Dressel L.L., Wolstencroft R.D., 1993, ApJ 414, 120

\ref
Tinsley B., 1980, Fund. Cosmic Physics 5, 287

\ref
Tomita A., Tomita Y., Saito M., 1996, PASJ 48, 285

\ref
Toomre, A., 1964, ApJ 139, 1217

\ref
Vila-Costa M.B., Edmunds M.G., 1992, MNRAS 259, 121

\ref 
van den Bergh S., Abraham R.G., Ellis R.S., et al., 1996, AJ 112, 359

\ref
Walsh J.R., Roy J-R., 1996, MNRAS, in press

\ref
Xu C., Helou G., 1996, ApJ 456, 152

\ref
Xu C., Klein U., Meinert D., et al., 1992, A\&A 257, 47

\ref
Young J.S., Xie S., Kenney J.D.P., Rice W.L., 1989, ApJS 70, 699

\ref
Zaritsky D., Kennicutt R.C., Huchra J.P., 1994, ApJ 420, 87

\endref

\bye